\documentclass{aa}
\usepackage{natbib}
\usepackage{graphicx}

\begin{document}
\def\teff{$T\rm_{eff }$}
\def\kms{$\mathrm {km s}^{-1}$}

\title{
Automatic 
abundance analysis of high resolution spectra
}

   \subtitle{}

\author{
P. \,Bonifacio \inst{1},
\and E. Caffau\inst{2}
          }

  \offprints{P. Bonifacio}

\institute{
Istituto Nazionale di Astrofisica --
Osservatorio Astronomico di Trieste, Via G.B.Tiepolo 11, 
I-34131 Trieste, Italy 
\and
Liceo Scientifico  Leonardo da Vinci, 12 Rue Sedillot, Paris, France
}

\authorrunning{Bonifacio \& Caffau}
\mail{bonifaci@ts.astro.it}

\titlerunning{Automatic abundance analysis 
}

\date{Received 18 September 2002 / Accepted 6 December 2002}

\abstract{
We describe an automatic procedure for determining abundances from
high resolution spectra. Such procedures are becoming increasingly
important as large amounts of data are delivered from 8m telescopes
and their high-multiplexing fiber facilities, such as FLAMES on ESO-VLT.
The present procedure is specifically targeted for the analysis
of spectra of giants in the Sgr dSph; however, the procedure may be,
in principle, tailored to analyse stars of any type.
Emphasis is placed on the algorithms and on the stability
of the method; the external accuracy rests, ultimately, 
on the reliability of the theoretical models (model-atmospheres,
synthetic spectra) used to interpret the data.
Comparison of the  results of the procedure with the
results of a traditional analysis for 12 Sgr giants shows that
abundances accurate at the level of 0.2 dex, comparable with
that of traditional analysis of the same spectra, 
may be derived in a fast and efficient way. 
Such automatic procedures are not meant to replace the traditional
abundance analysis, but as an aid to extract rapidly a good deal
of the information contained in the spectra.
}
\maketitle{}           

\begin{figure*}
   \centering
\resizebox{\hsize}{!}{\includegraphics[clip=true]{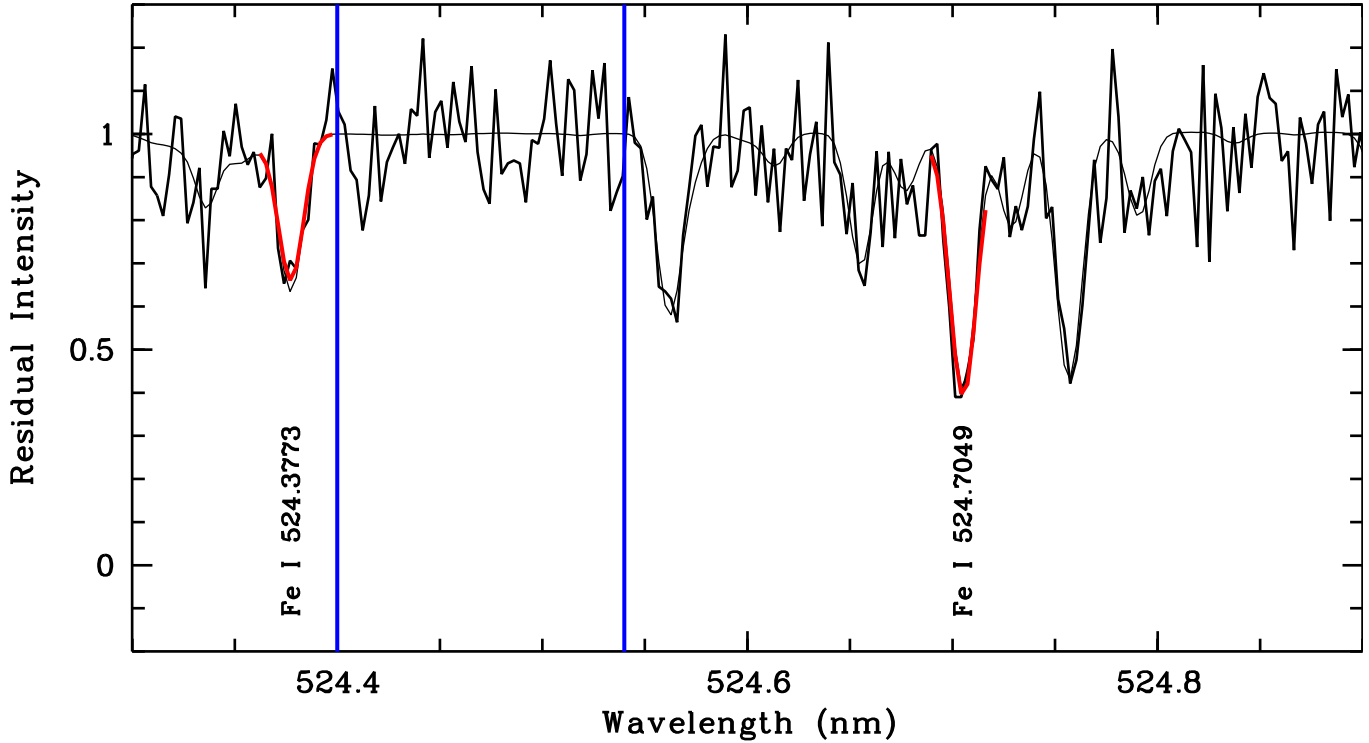}}
\caption{Example of fit to a simulated noisy spectrum.
The thin line is a synthetic spectrum corresponding to 
T$_{eff}=5000$ K, log g = 2.50, [Fe/H]=--0.5, [$\alpha$/Fe]=+0.4, $\xi=1.0
{\rm kms^{-1}}$ and 
pseudo-normalized as described in the text.
The thicker line is the same spectrum  
in which noise has been injected so that S/N = 10 and
pseudo-normalized {\em after} the injection of noise.
The two vertical lines delimit a pseudo-continuum window.
The two thick lines are the fitted spectra of two
Fe\,\textsc{i} features.}
\label{example}
\end{figure*}

\begin{figure}
   \centering
\resizebox{\hsize}{!}{\includegraphics[clip=true]{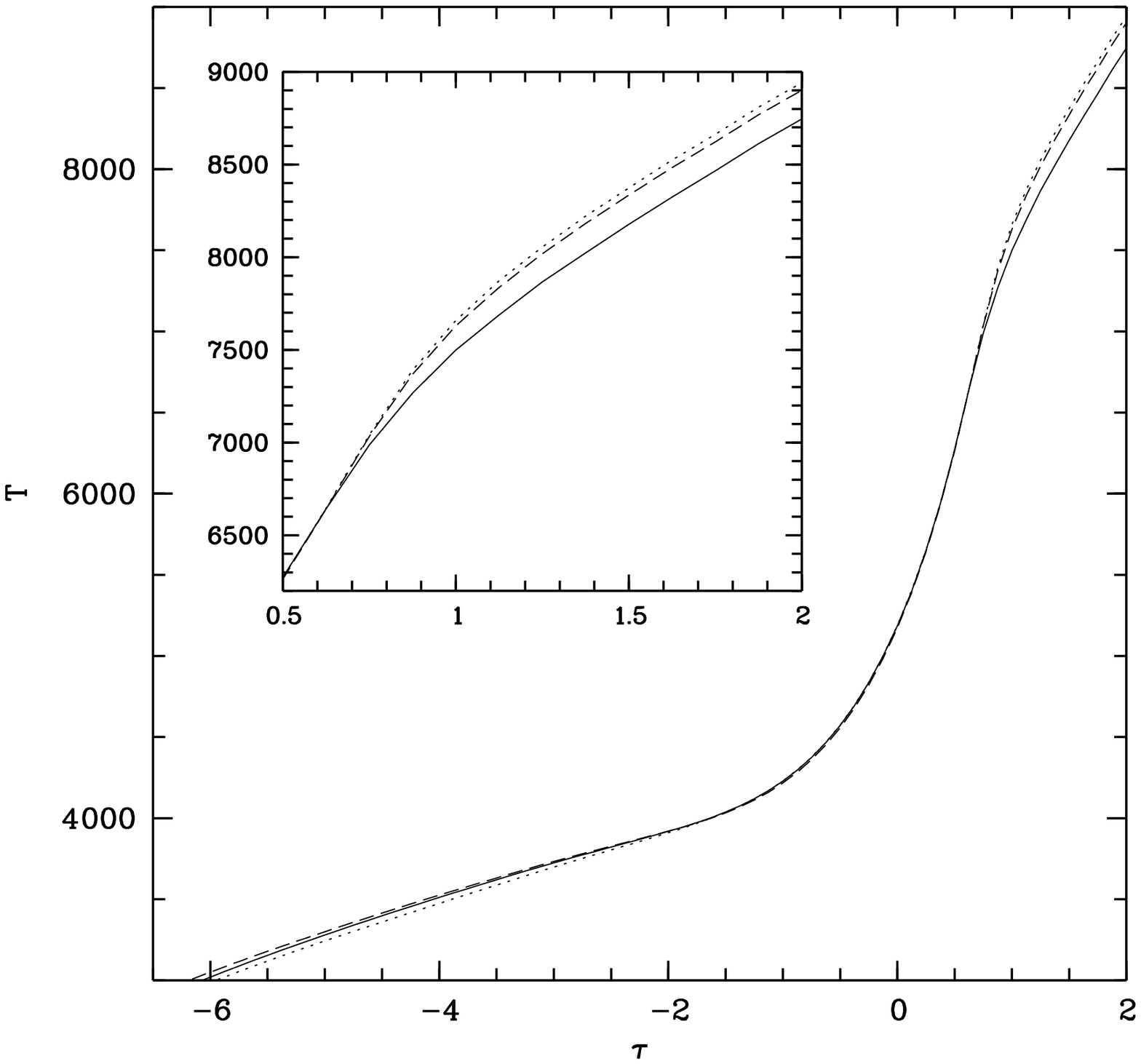}}
\caption{Optical depth as a function of temperature
for three models with T=4750 K and log g=2.5.
The solid line corresponds to a model with [M/H]=--0.5 and
[$\alpha $/Fe]=0.0 and the dashed line to one with same
[M/H] and [$\alpha $/Fe]=0.4;
the dotted line corresponds to a model with [M/H]=--0.19
and [$\alpha $/Fe]=0.0. In the inset is shown a zoom 
of the temperature structure of the models at large
optical depths. 
A difference of about 60 K between the model with
[M/H]=--0.5, [$\alpha /Fe$]=0.4
and the model with [M/H]=--0.19, [$\alpha /Fe$]=0.0 
can be appreciated.}
\label{zeta}
\end{figure}

\section{Introduction}

In recent years considerable attention has been 
devoted to full or partial automation of the
process of abundance determination
from high resolution spectra 
\citep{Katz98,KatzT,EN02,EN02b}.
One of the main thrusts behind these attempts is the increasingly
large amount of data delivered by modern instrumentation.
It is well known  that astronomical archives
are full of good quality high resolution spectra which have not
been analysed, or only partially analysed, due to lack of manpower.
In spite of the complexity of data reduction, this is not
the bottleneck, thanks to the
efficient software and instrument-dedicated pipelines that are available.
The real bottleneck is the data analysis which, for abundances, is still
done more or less in the same way as twenty years ago.
The situation is going to become even more critical as 
high-multiplex 
instruments, such as FLAMES on ESO-VLT \citep{Pasquini}, become fully
operative.

In the last years we have devoted special interest to the
determination of abundances in giants of the Sgr dSph
\citep{B99,bonivlt,B00}. 
Such an astrophysical problem is ideal for the capabilities of
FLAMES. It has also the advantage that two of the
key parameters in the interpretation of stellar spectra,
effective temperature and surface gravity, may be conveniently
constrained. If we pick stars of roughly the same apparent magnitude,
if they belong to Sgr, they will have the same luminosity and 
therefore surface gravity. At a given luminosity
the RGB of Sgr spans a limited range in effective temperature,
in fact less than 1000 K, and in the case of our sample 
less than 250 K. 
We had already attempted to develop a procedure to analyse
the low resolution spectra which we obtained from
EMMI-NTT \citep{B99,bonivlt}, however the procedure, based on
spectral indices, provided 
unsatisfactory results, essentially because of the combination
of low resolution and low S/N ratios. Furthermore it was apparent
that at low resolution we had no handle on the microturbulent
velocity, and since the abundances relied only on strong lines
(the only ones available at low resolution, whichever the S/N ratio),
the result was strongly dependent on the unknown microturbulent
velocity.
Having in mind the future use of FLAMES and having available
a few UVES spectra, obtained in slit-mode, we decided to concentrate
on a procedure capable of analysing spectra from UVES and Giraffe.
We have developed a procedure which is very efficient and stable,
at least on the real UVES spectra. In spite of the fact
that it is highly targeted (it will deal only with stars of
a given luminosity and a limited range of effective temperatures),
it has been written in such a way that it may be modified
to deal with other types of stars. 
In this paper we describe the procedure placing our emphasis
on the algorithms and on the stability of the method.
The procedure rests on synthetic spectra computed from
1D LTE model atmospheres. In this paper we do not question
the reliability of the input synthetic spectra, since it is
straightforward to replace the existing grid with a better one,
when available.

\begin{table} 
\caption{Algorithm meta-code}
\label{algorithm}
\begin{center}
\begin{tabular}{lll}
\hline
\\ 
\\
BEGIN &\\
  & $\xi$;\\
  &&  pseudo-normalize;\\
  &&  find [Fe/H] for each FeI feature;\\
  &&  find slope A(FeI) vs.  EW;\\
  &&  if slope$<$ threshold, go to $\alpha$;\\
  &&  find a $\xi$ to make slope smaller;\\
  & goto  $\xi$;\\
  & $\alpha$;\\
  &&  find  [Mg/H] for each Mg I  feature;\\
  &&  find  [Ca/H] for each Ca I feature;\\
   && [$\alpha$/Fe] = mean of [Mg/Fe] and [Ca/Fe];\\
  && if  $\Delta$[$\alpha$/Fe] $<$ threshold1 goto END;\\
  & goto $\xi$;\\
END   \\

\\
\hline
\\
\end{tabular}
\end{center}
\end{table}

\begin{figure}
\centering
\resizebox{\hsize}{!}{\includegraphics[clip=true]{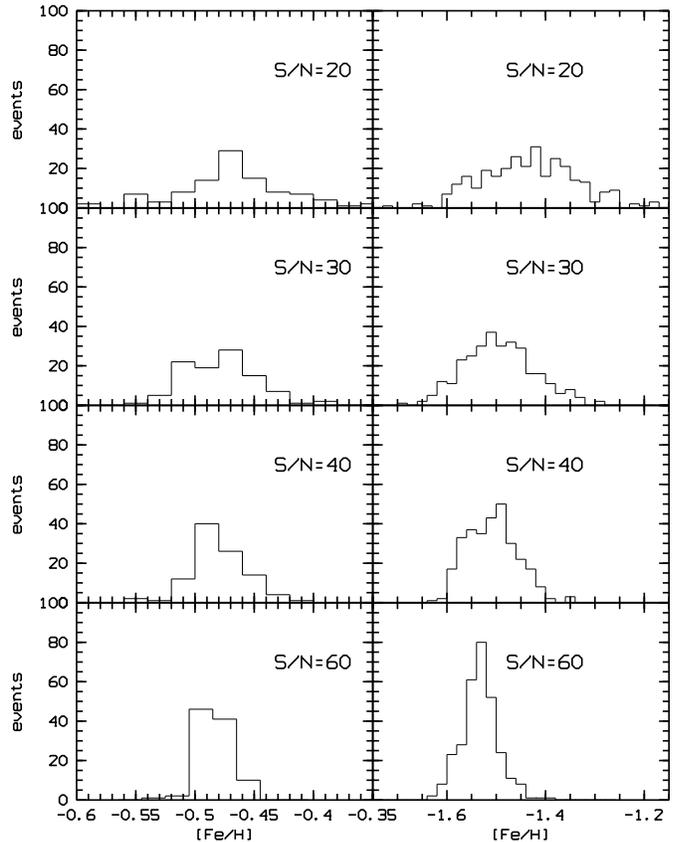}}
\caption{Histogram of the derived [Fe/H] for Monte Carlo simulations
with different S/N ratios and a resolving power of
7 kms$^{-1}$}
\label{histo}
\end{figure}

\begin{figure}
\centering
\resizebox{\hsize}{!}{\includegraphics[clip=true]{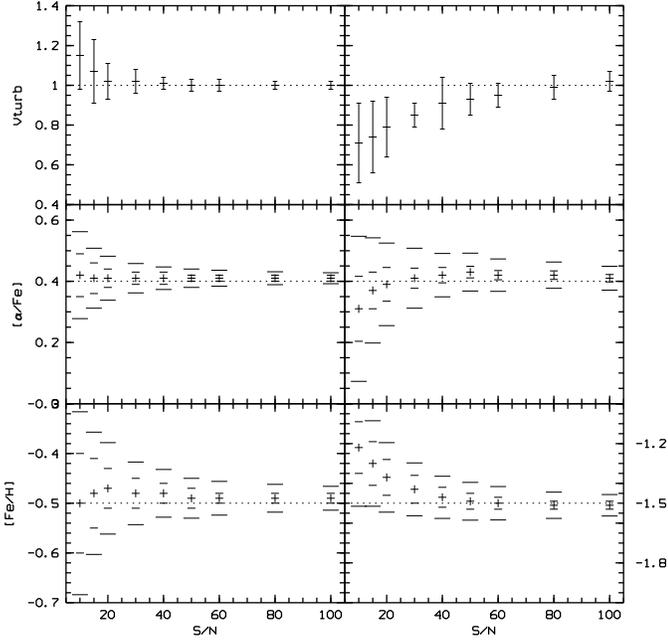}}
\caption{
Monte Carlo simulations corresponding
to a resolving power of
7 kms$^{-1}$:
[Fe/H], [$\alpha$/Fe] and microturbulence velocity as a function
of S/N for input spectrum at $\xi =1.$Km/s, [$\alpha $/Fe]=0.4 and
[Fe/H]=--0.5 on the left panel and [Fe/H]=--1.5 on the right panel.
For both the [Fe/H] vs S/N plots the smaller error bars correspond to
Monte Carlo simulation dispersion, the bigger one to the [Fe/H]
dispersion over the all Fe\,\textsc{i} features.}
\label{snplot}
\end{figure}

\begin{figure*}
   \centering
\resizebox{16cm}{!}{\includegraphics[clip=true]{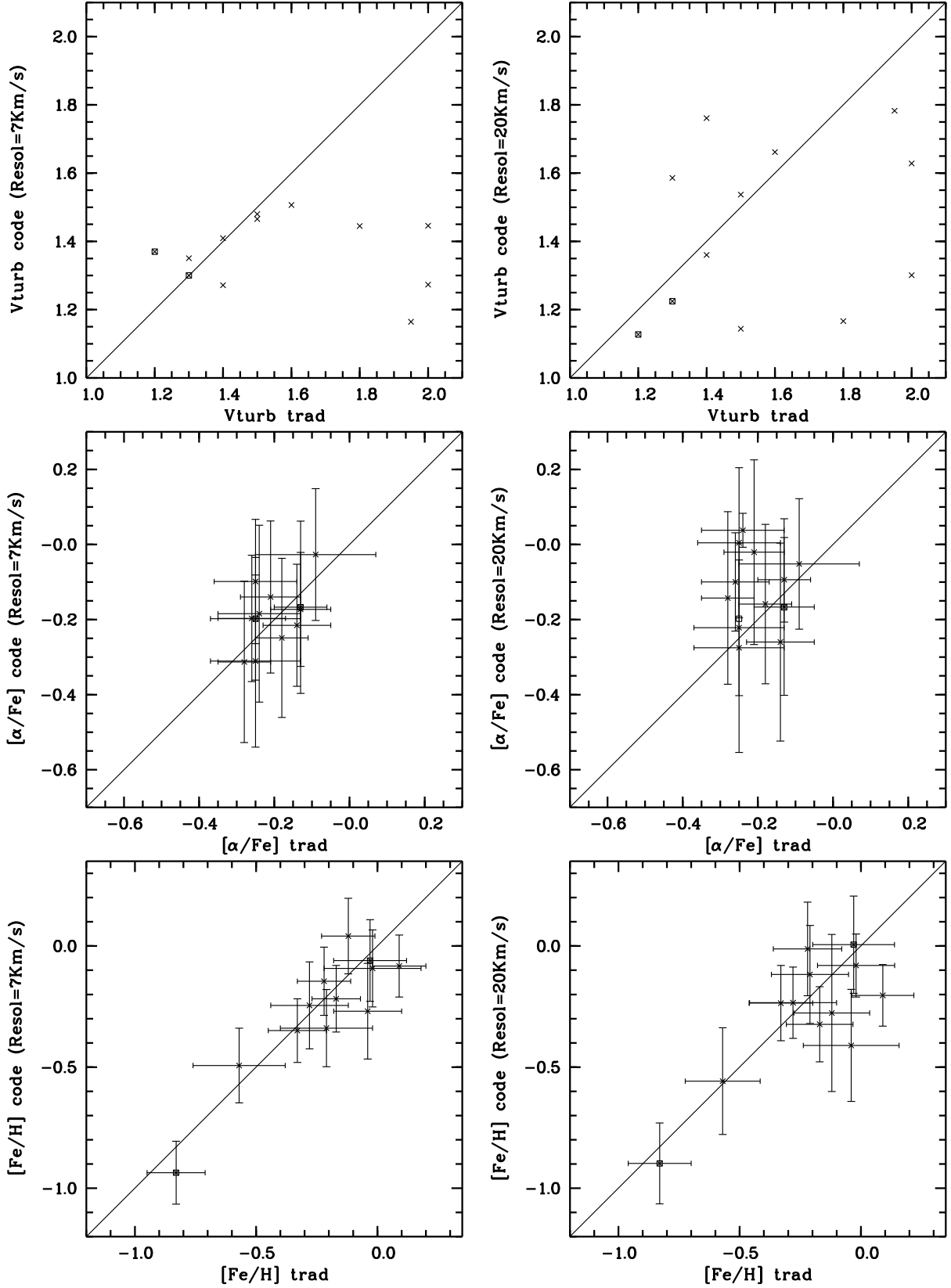}}
\caption{Comparison of [Fe/H], [$\alpha$/Fe] and
$\xi$ from the traditional analysis and 
the automatic analysis for the 12 stars
of Sgr observed with UVES. 
The left panels display  the results from the UVES
spectra, i.e. at a resolving power of $\sim 7\rm kms^{-1}$;
the right panels display  the results 
from the UVES spectra convolved with a gaussian profile,
obtaining a resolving power of $\sim 20 \rm kms^{-1}$.
The two points, in each of the eight panels,  shown as
crossed squares, identify the two stars for which the
log g adopted in the traditional analysis is not 2.5.}
\label{conf}
\end{figure*} 

\begin{figure}
   \centering
\resizebox{8.5cm}{!}{\includegraphics[clip=true]{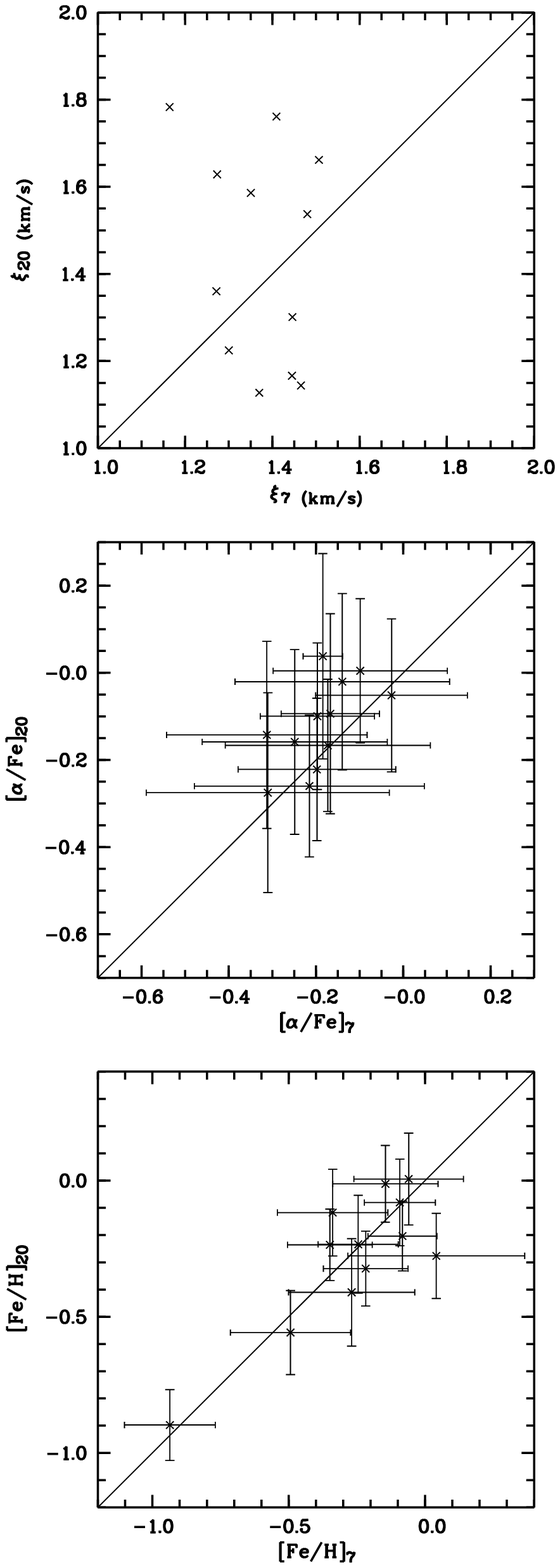}}
\caption{Comparison of [Fe/H], [$\alpha $/Fe] and
$\xi$ obtained for  the 12 Sgr giants
observed with UVES directly from the UVES spectra
(resolving power $\sim \rm 7 kms^{-1}$)
and UVES spectra broadened to a resolving power
of $\sim \rm  20kms^{-1} $.}
\label{br7_20}
\end{figure}

\section{Algorithms }

The algorithms employed by our code reproduce 
those usually employed in abundance analysis
of stellar spectra.
The whole procedure rests on the comparison between 
computed and observed spectra.
 We assume that the observed spectra have been fully 
reduced and shifted to rest-wavelength.
For the Giraffe spectra it is expected that
the instrument pipeline will provide 
radial velocities through cross-correlation.
The basic algorithm of our code is summarized as
meta-code in Table \ref{algorithm}.

\subsection{Pseudo-normalization}

The first step is the normalization of the spectra:
this is done by defining a set of continuum windows.
In each window the continuum is defined
as a quasi-median: if there are $N$ pixels
in the window the pixels are sorted by
number of counts; let $n= [3N/5]$, 
the continuum is defined as the
number of counts  in the $n-$th pixel.
This definition was chosen because extensive 
Monte Carlo simulations showed that it performed
better than the  median.
 The point in the (wavelength, counts) plane
defined by the central wavelength and this quasi-median 
of the continuum window is considered as a continuum, 
or rather
pseudo-continuum, point. 
A spline is then passed through 
these  points  and the spectrum pseudo-normalized
dividing it, pixel by pixel, by this spline. 
It is impossible to
define true continuum windows, each one is affected
by weak absorptions, which are larger the lower the 
effective temperature and the higher the metallicity.
 In Fig. \ref{example} we show an example
of a typical continuum window and of
pseudo-normalized spectra in a simulation with a synthetic spectrum.
In order to make a meaningful comparison the computed spectra 
are normalized (or pseudo-normalized, to be rigorous)
exactly in the same manner as observed spectra.
One could fear that
the synthetic and the observed spectrum
are not normalized exactly in the same way,
since the observed spectrum may have
a complicated shape due to
variations of the instrumental efficiency (CCD QE, blaze function and, in
the case of Giraffe, transmission of the order-separating filters).
To avoid these problems,
the continuum windows on the one hand
must be  small enough that 
the instrument response function $\cal R$ 
is  constant in the window.
In this case  the pseudo-normalized spectrum 
in the window is $ F_\lambda{\cal R}/<F_\lambda{\cal R}>\sim 
F_\lambda/<F_\lambda>$,
where $F_\lambda$ denotes the incoming photon flux and 
the angle brackets denote the quasi-median.
It is thus approximately independent of 
the instrument response function.
On the other hand there must be enough pseudo-continuum windows
to capture the variations of the
instrument response function.
It is clear that the width and number of
pseudo-continuum windows must be chosen also to satisfy
these requirements and is therefore instrument-dependent.
In the case of the UVES data the instrument response
function is very smoothly varying.  Our chosen
continuum windows have widths $w$ in the range $0.05{\rm nm} \la w \la 
0.4{\rm nm}$
and the distance $d$ between windows is in the range
$0.1{\rm nm}\la d \la 3.0{\rm nm} $ with an average $d\sim 1$nm.
It is possible that, when we have at our disposal true Giraffe
data, we shall have to revise the continuum windows.
However one must keep in mind that, in the process of data reduction
prior to the analysis, one tries to remove
the instrument response function as well as possible (``rectify''
the spectra), and one is left with a trend which is 
smooth and well approximated by a spline.
With UVES spectra this is very well achieved over intervals
of $\sim $ 20nm such as we are considering. 
Inspection of Fig. \ref{example} suggests that
the correct position of the pseudo-continuum is recovered
even at very low S/N levels.

Since our work is specifically targeted to the 
study of giants in the Sagittarius dwarf Spheroidal
using the FLAMES facility,
our program presently runs assuming that two
spectral ranges, corresponding to high resolution settings 
HR09 and HR14 of the Giraffe spectrograph,  are available.
These were chosen
in order to give information on Fe,
the $\alpha $ process elements Ca and Mg, as well
as Si, Ti, the iron peak elements Sc, V and Ni, and
the n-capture elements Ba and La,
thus providing information on three distinct
nucleosynthetic channels.	
In the present paper we focus on the
Fe, Mg and Ca, however extension to all the other
available elements is under way and some results are presented
in section 5.2.

\subsection{Selected spectral features}

A number of spectral regions are selected, each one
is dominated by one line of a given ion. 
 In Fig. \ref{example} two typical Fe\,\textsc{i}
features used are labeled and the fitted profiles are shown.
In the following
we refer to these as {\em features} and not {\em lines},
since, although dominated by a line, they are affected
by several blending lines whose relative importance
changes with atmospheric parameters.
These windows have been optimized for a resolution
of 7 $\rm kms^{-1}$, relevant to the UVES data on which
the procedure has been tested. At the lower resolution 
of Giraffe these windows change slightly.
For each range a comparison between the \citet{Kursolat}
solar atlas and a
synthetic spectrum computed from the Kurucz solar  
model \citep{k93} has been done in order to
check   the considered features.
We have rejected some features,
which were very poorly reproduced by the synthetic spectrum.
However, for  some  of the features 
which we retained, the discrepancies between 
synthetic and observed spectrum are not negligible.
Had we rejected all of these, we would have been left
with too few features to base our analysis on.
A possibility would have been to modify the atomic
data used for computing our synthetic spectra,
in order to force a good fit of the solar spectrum,
as done, for instance by \citet{chavez}.
However since the stars in Sgr, which
we want to study,
have temperature,
gravity and, perhaps, metallicity very different from the Sun,
there is no guarantee that changes to the atomic data
made to obtain a better fit in the Sun will ensure
also a better fit in our stars.
Moreover it is not obvious how much of the disagreement
is due to the atomic data and how much to the temperature structure
of the model atmosphere \citep[see for example][]{sauval}.
Therefore rather than introducing {\em ad hoc}
changes, whose effect is difficult to predict, we prefer
to keep features for which our capability to reproduce
the solar spectrum is less than optimal and validate
our procedure by comparing the results with that of
a traditional abundance analysis, based on different lines.

\subsection{Fitting observed features}

The comparison of observed and synthetic
spectra relies  on least-square fitting.
For each
region 
we define $\displaystyle\chi ^2=\sum \left({O_i - S_i\over\sigma}\right)^2$
where $O_i$ is the observed spectrum for the $i-th$ pixel,
$S_i$ is the synthetic spectrum
and $\sigma$, assumed equal for all pixels, is defined
as $ \displaystyle 
\left[{1\over 2}\left({<C_l>\over\sigma_l}+
{<C_r>\over\sigma_r}\right)\right]^{-1},
 $ where $ <C_l>$ and $<C_r> $ are the mean number
of counts in the nearest continuum windows to the left and
to the right of the given feature; $\sigma_l$ and
$ \sigma_r $ are the standard deviations in the same.
At each step in the code the fitting of any feature
has two parameters: the abundance and
a small shift
in central wavelength of up to one
resolution element.
Since the fit is done on each line separately this
is specific to each spectral window and not
a global shift.
The minimum $\chi^2$ is sought numerically using
MINUIT \citep{cern}.
Two comments are in order here:
\begin{enumerate}
\item  The theorems on the statistical properties of
       the $\chi^2$ are not applicable in the case
       of real spectra.  If the spectra have been rebinned to
       a constant wavelength step, as is usually done, the
       neighbouring pixels are obviously correlated.
       This is however true also for non-rebinned spectra;
       since the slit width normally projects onto several
       pixels, thus adjacent pixels belonging to the same 
       resolution element are correlated.
\item  The assumption that all pixels have the same $\sigma$
       is acceptable for weak lines but it clearly breaks
       down for a strong absorption line, whose central residual
       intensity is near to zero.
\end{enumerate}

In spite of this, our Monte Carlo simulations show that
$\chi^2$ fitting may be used  to determine the correct
values of the fitting parameters (in our case abundances)
and the error on the fit may be conveniently estimated
using Monte Carlo simulations.
On the other hand the "goodness of fit" derived from
the $\chi^2$ distribution may have little meaning in
an absolute sense.
However we need some way to discriminate between ``good''
and ``bad'' fits in order to reject the latter.
Our pragmatic solution is to use the formal
``goodness of fit'' but to consider the
minimum probability to accept a fit to be a free parameter
which must be tuned to the actual data to be analysed, based
on the results of Monte Carlo simulations and, when available,
on the results for stars of known abundance.

\subsection{Fitting parameters}

Given that computing synthetic spectra is time consuming,
these are pre-tabulated in a 
spectrum library, and linear interpolation is used when 
necessary.
We assume that the effective temperature and the surface
gravity of the star are known.
The Sgr giants, we are interested in, have more or less the
same gravity and effective temperature may be derived 
e.g. by a suitable colour such as (V--I). We shall
estimate the errors due to incorrect choice of temperature
or gravity.
As detailed in the next section, our synthetic grid
is based on standard, flux constant, plane parallel 
model atmospheres computed with the ATLAS 9  code.
Therefore the atmospheric parameters to be estimated 
are: metallicity, which we define for practical reasons
to be [Fe/H]\footnote{We adopt throughout the paper the usual
notation [X/H]= log [N(X)/N(H)] +12 .}, the microturbulent
velocity and the $\alpha$ enhancement.
The latter parameter is often neglected in abundance
analysis. In fact it is customary to neglect the effect
of chemical composition on the atmospheric structure.
While this is acceptable for trace elements, such as Li, it is not
acceptable for the $\alpha$ elements, at least for the
most metal-rich models. 

As an example we show in Fig. \ref{zeta}  the run of 
temperature vs Rosseland optical depth for three models
with the same effective temperature, gravity and microturbulence. 
The solid line and the dashed line correspond
to [M/H]$=-0.5$ and [$\alpha$/Fe]=0.0 and +0.4
respectively.
The differences at large optical depths, where weak lines
are formed, are as large as 200 K as shown in the inset.
Apart from the different temperature structure of the
models the abundance of $\alpha$ elements has a direct
impact on the pseudo-continuum, especially in the region
from 510 nm to 525 nm, which is filled by weak MgH lines, 
which affect also the continuum windows.
In reality one should not label the models by their
value of [Fe/H], but rather by their Z.
The   model with 
[M/H]=--0.5 and  [$\alpha $/Fe]=0.4 has Z=0.0061;
the model with the same [M/H], but
[$\alpha $/Fe]=0.0, has 
Z=0.0126.
Therefore a 
model with [$\alpha $/Fe]=0.4 and [M/H]=--0.5 has about the
same Z as a model with [$\alpha $/Fe]=0.0 but [M/H]=--0.19
(to be precise a model with [M/H]=--0.186 has Z=0.0126).
We calculated such a  model using an appropriate ODF
and found that, although the Z is the same, the
temperature structure of the two models
still differs in depth by  about 60 K as shown in Fig. \ref{zeta}.
Therefore, although Z is a better description of the
overall properties of the model than [Fe/H] or [M/H] and [$\alpha$/Fe]
the details still depend on the precise chemical composition.
Moreover in order to know Z one must know the abundances of all
elements, or, at least C, N and O, which, in general, are quite difficult
to obtain. 
On the contrary [Fe/H], [Mg/H] and [Ca/H] are relatively 
easy to obtain, it is thus not surprising that
for practical reasons, when  
studying  stellar  spectra, the chemical composition 
is more conveniently described in terms of  [Fe/H] and [$\alpha$/Fe]
rather than Z.

\subsection{Iterative steps}

 As outlined in the meta-code in Table \ref{algorithm} 
after the normalization of the spectra the next step is to determine
the metallicity by fitting the Fe\,\textsc{i} features,  
in the above
defined sense. 
For each feature the equivalent width (EW) is determined
by trapezoidal integration of the fitted  synthetic profile.
Then the slope in the plane (log (abundance), log EW)
is found. 
The computation is repeated for at least two
values of
the microturbulence $\xi$.
The  Fe\,\textsc{i}  abundances are re-determined with the new
microturbulence and the process is iterated until
the residual slope is below a chosen value 
(experience has shown that a value
like 0.001 log[EW(nm)]/dex is a good choice in practical cases).
 A possible concern is that in this step we
are treating our features as if they were isolated lines.
Suppose a feature were composed by many weak  Fe\,\textsc{i}
lines. It would be insensitive to microturbulence,
since all the lines are on the linear part of the 
curve of growth. Yet its equivalent width could be
large. The presence of several
such features would force a small microturbulence
velocity, when  a higher one would be required.
In practice one has to avoid such cases. We selected
our features so that they behave, with respect to the
microturbulent velocity, in the same way as isolated 
lines would do.  
Having fixed the microturbulence, the $\alpha$ to iron
ratio is determined by fitting Mg and Ca lines. 
Once a value of [$\alpha$/Fe] is found, the process is iterated,
 as shown in Table \ref{algorithm},
starting from Fe abundances, through 
determination of microturbulent velocity and a new value
of  [$\alpha$/Fe], until the change in  [$\alpha$/Fe] is  smaller
than a chosen value (0.01 dex has been used in practical cases).
Readers familiar with abundance analysis will recognize that
this whole procedure is identical to what is used in
traditional ``fine abundance analysis'', except that in the latter
 isolated lines, rather than features formed by line blends are
used. Moreover
often effective temperature is adjusted to satisfy
excitation equilibria and surface gravity to satisfy
ionization equilibria. 
It would be in fact straightforward to implement
these in our automatic scheme, however we reckon these
adjustments are not important in the case of Sgr giants and
prefer to simply estimate the error in abundances due to
an incorrectly chosen temperature and gravity.
Once metallicity, microturbulence and  [$\alpha$/Fe] 
are fixed, the procedure can continue by determining the abundances
of other available elements by fitting the relevant
spectral features.

\subsection{Stars with enhanced C and N abundances}
Although so far
none of the Sgr giants has been found to have
enhanced CN bands, such stars are known to exist 
among field and globular cluster giants.
The chosen spectral
regions are characterized by the presence
of many CN lines which are very weak
in the case of solar or solar-scaled chemical
compositions, however they become strong
in the presence of
very large enhancements of C and N.
Such stars, if present, cannot be dealt with
by the present procedure.
We therefore identified a number of
windows to be used as ``CN alert''.
These are regions characterized by very weak CN and/or CH
lines,
in the case of solar or solar-scaled chemical composition,
which become appreciable  
in the case of conspicuous CN enhancement.
We compute the absolute difference between the
observed spectrum and the ``best fitting''
spectrum and use this and
a user-defined threshold
to set a ``suspect CN'' flag, which should 
prompt for further inspection of the spectrum.
Tests with synthetic spectra have shown that we can 
 discriminate against CN enhancements of
[C/Fe] $\ge +0.4$ and [N/Fe]$\ge +0.4$.
The C and N enhancements observed in C-rich N-rich stars
are often much larger than this.

\section{Grid of synthetic spectra}

The synthetic spectra  employed
have been computed 
from LTE model atmospheres 
computed with the ATLAS code,
using the SYNTHE
code \citep{k93}.  
The spectra cover the ranges
corresponding to settings HR09 and HR14 of the Giraffe spectrograph.
The grid covers 5 effective temperatures (4500 K, 4750 K, 
5000 K, 5250 K, 5500 K), 5 metallicities ([M/H]=--2.0, --1.5, --1.0, --0.5, 0.0),
two values of $\alpha$ enhancement (0.0, +0.4) and 
three microturbulent velocities ($\xi=1.0,~2.0,~3.0$ kms$^{-1}$)
for a single value of log g = 2.5, thus it presently
consists of 150 spectra for each of the above ranges.
In the future the grid may be extended.

In the course of the fitting procedure the spectra are convolved
with an instrumental profile and rebinned in wavelength in order
to have the same step as the observed spectra.
So far the procedure has been tested only on UVES data 
(R$\sim$ 40000) and
on UVES data degraded to the expected resolution of Giraffe
(R$\sim 15000$).  
The referee, P. North, pointed out that the high resolution mode of Giraffe,
tested during the commissioning of the instrument,
is higher than expected and in fact around 20000.
To verify the impact of the increased resolution we performed
a small set of Monte Carlo simulations also at this resolution,
as detailed in the next section.

\section{Monte Carlo simulations}

\subsection{Stability of the method}

In order to estimate the stability  of the method we resorted to
Monte Carlo simulations. 
We took two  synthetic spectra, 
with T = 5000 K, [$\alpha$/M]=+0.4, $\xi=1.0 $ kms$^{-1}$
and metallicities  [M/H]=$-0.5$ and $-1.5$.
These were 
convolved with a Gaussian instrumental profile of 7 kms$^{-1}$
(to simulate a UVES spectrum) of  20 kms$^{-1}$ and
a few also of 15   kms$^{-1}$
(to simulate a Giraffe spectrum).
We injected Gaussian noise in order to reach
various levels of 
S/N  ratio  (10, 15, 20, 30, 40, 50, 60, 80, 100) 
and ran the procedure  for 300 MC samples in each case.
For the resolution of 15   kms$^{-1}$
only S/N ratios of 20, 30, 40, 60, 80 were simulated.
We verified that after about 300 samples the dispersions stabilized
so as to make it unnecessary to run a larger number of samples.
 The histograms of the derived metallicity in the case
of a resolving power of 7 kms$^{-1}$
are shown in figure
\ref{histo}. As expected the dispersion increases quite
rapidly with decreasing S/N, however the dispersion at S/N = 20
is only about 0.05 dex.
This is an ``internal'' error
of the method and does not take into
account systematic errors due to shortcomings of the 
synthetic grid 
(errors in the atomic data, inadequacy of model-atmospheres...)
nor due to an incorrect choice of temperature or gravity.  
Furthermore the assumption of Gaussian noise
is seldom a satisfactory one for real spectra, which are affected
by cosmic ray hits, sky emission lines, and other sources
of non Gaussian noise.
Therefore our Monte Carlo simulations provide a check
on the stability of the method, rather than an estimate
of the accuracy of our procedure.
Moreover in a practical case it is not straightforward how
the S/N ratio should be estimated. Instead the
dispersion of abundances derived from different features is
a readily available parameter, which we suggest to use as
error estimate. The Monte Carlo simulations show that this
dispersion is always within a factor two of the 
Monte Carlo error, thus making it a suitable, although 
somewhat conservative,
error estimate (see fig.\ref{snplot}).

It is interesting to notice in Figures \ref{histo} and
\ref{snplot}
how going to lower S/N ratios, the estimated metallicity
shifts systematically to-wards higher values, although 
by a relatively small amount: for the simulations with
[Fe/H]=--1.5 going from S/N=60 to S/N=20 the estimated
metallicity increases only by about 0.1 dex.
For the simulations with [Fe/H]=--0.5  
the increase is negligible.

We estimated the error introduced by an incorrect choice
of temperature and gravity by feeding to the procedure
noise injected synthetic spectra with \teff = $\pm 100$ K 
and log g = $\pm 0.5$ dex .
The corresponding changes in metallicity and
$\alpha$ enhancement are $\pm 0.1$ dex and $\pm 0.04$ dex
respectively for a 100 K  change in \teff ~and
$\pm 0.04$ dex and $\mp 0.10$ dex 
respectively for a change
of $\mp 0.5$ dex in log g.

 The Monte Carlo simulations also provide a
way of assessing the effect of the different resolutions.
We ran Monte Carlo simulations on a noisy synthetic
spectrum with [Fe/H]=--0.5,  [$\alpha$/Fe]=+0.4, 
$\xi=1.0\rm kms^{-1}$ and S/N=20 with
resolving power 7, 15 and 20 $\rm kms^{-1}$ .
The results, displayed in Table \ref{confresolmc}, 
imply that the higher resolution
attains a smaller dispersion, but the mean values
are hardly changed.

\begin{table} 
\caption{Results of automatic analysis at different resolutions
for S/N=20}
\label{confresolmc}
\begin{center}
\begin{tabular}{ccccccccc}
\hline
\\
R & \multispan3{\hfill [Fe/H]\hfill} & \multispan3{\hfill [$\alpha$/Fe]\hfill } & $\xi$ & $\sigma _{\xi}$ \\
$10^3$ & & $\sigma$ &$\sigma _{mc}$ &  & $\sigma$ &$\sigma _{mc}$ & \multispan2{\hfill \kms \hfill} \\
\\
\hline
\\
40 & --0.47 & 0.09 & 0.04 & 0.41 & 0.07 & 0.03 & 1.0 & 0.1\\
20 & --0.47 & 0.15 & 0.08 & 0.40 & 0.11 & 0.06 & 1.0 & 0.2\\
15 & --0.47 & 0.18 & 0.10 & 0.40 & 0.13 & 0.06 & 1.1 & 0.2\\
\\
\hline
\end{tabular}
\end{center}
\end{table}

\subsection{Discrimination against CN rich stars}

The parameter which controls the ``CN-rich'' flag
is a pseudo-equivalent width per/pixel.
After some trials we decided that for our spectra
1.0 pm  (i.e. $10^{-12}$m) is a suitable value.
At a different resolution, like that of Giraffe for example,
this value has to be slightly modified.
Using the Monte Carlo method 
and our choice of ``CN alert'' and ranges
we obtain that every input ``CN-rich'' spectrum is detected
as  ``CN-rich''   both at S/N=30, and S/N=20.
On the other hand a ``CN-normal''  star is never flagged as ``CN-rich''.

\section{Application to spectra of giant stars in the Sgr dwarf galaxy}

We applied our procedure to UVES spectra of 12 giants of 
the Sgr dSph.  Two have been analyzed in the traditional
way by \citet{B00}, the remaining 10 have been analyzed
by \citet{B02}.
For the 10 stars analyzed by
\citet{B02} the radial velocities were
measured by cross-correlation with synthetic
spectra while \citet{B00}
measured the line positions of $\sim 60$ lines for each star.
The observed spectra have been  shifted to rest wavelength. 

\subsection{Metallicity,  $\alpha$ to iron ratio and microturbulence}
The results of this comparison is summarized in Table \ref{codetrad}
and is plotted in Fig. \ref{conf}.
It may be seen that the metallicity and the $\alpha$ enhancement
derived from our automatic procedure are consistent
at $1 \sigma$ with the results of a traditional analysis,
when the dispersion of results of individual features is taken as
error estimate. 
This in spite of the fact
that in the traditional analysis in two cases
the gravity has been changed to
satisfy the iron ionization equilibrium, while no such thing
is possible in the present version of our procedure.
The mean metallicity difference in the sense
(code -- traditional) is $-0.04$ dex with
a standard deviation of  0.11 dex.
The mean difference in  [$\alpha$/Fe]  
is +0.013 dex with a standard deviation of
0.074 dex.
The mean difference in $\xi$ is 
 $-0.21$ \kms with a standard deviation of
0.32 \kms.
 A possible cause of concern, looking at the top panel
of Fig. \ref{conf}, is that there appears to be no correlation
between the $\xi$ measured by the code and the traditional
abundance analysis. 
In fact what happens is that for eight of the stars there
is a tight correlation, whereas for the remaining four 
there is no significant correlation, as can be inferred
by computing Kendall's $\tau$ for the two subsets.
The reason for which $\xi$ is poorly determined in some cases
is that, in general, there are few weak Fe\,\textsc{i} 
features. At low S/N ratios it may happen that 
most of them are rejected because the goodness-of-fit is
too small. In these conditions $\xi$ is essentially undetermined.
The comparison between automatic and traditional analysis  
suggests that the accuracy on $\xi$ is of the order of 0.3 $\rm kms^{-1}$,
which  implies an error in abundance
of the order of 0.2 dex, for saturated lines.
Since this is within the claimed accuracy of the method we 
believe it is not a serious problem.
Furthermore we point out that several of the Ca\,\textsc{i} 
lines used for the determination of [$\alpha$/Fe] are
saturated and the good agreement of these ratios between
automatic and traditional analysis supports the above claim.

\begin{table*} 
\caption{Results of automatic and traditional analysis}
\label{codetrad}
\begin{center}
\begin{tabular}{lcccllllllll}
\hline
\\ 
Star$^a$ & S/N &$\rm T_{eff}$ & \multispan2{gravity} &  \multispan2{[Fe/H]}  &  \multispan2{[$\alpha$/Fe]} &  
\multispan2{$\xi$ (km/s)} \\
\\
& @530nm & & code & trad & code & trad & code & trad & code & trad  \\
\\
\hline
\\
879$^b$ & 17 & 4891. & 2.50 & 2.50 & $-0.25\pm 0.18$ & $-0.28\pm 0.16$ &$-0.17\pm 0.15 $ & $-0.13\pm 0.08$ & 1.3 & 1.4  \\ 
772$^c$ & 18 & 4891. & 2.50 & 2.50 & $-0.34\pm 0.16$ & $-0.21\pm 0.19$ &$-0.31\pm 0.23 $ & $-0.25\pm 0.12$ & 1.5 & 1.5  \\ 
628     & 37 & 4904. & 2.50 & 2.50 & $-0.15\pm 0.14$ & $-0.22\pm 0.11$ &$-0.22\pm 0.16 $ & $-0.14\pm 0.09$ & 1.5 & 2.0  \\ 
656     & 24 & 5017. & 2.50 & 2.50 & $-0.22\pm 0.14$ & $-0.17\pm 0.10$ &$-0.14\pm 0.20 $ & $-0.21\pm 0.08$ & 1.5 & 1.6  \\ 
716     & 36 & 4967. & 2.50 & 2.50 & $+0.04\pm 0.16$ & $-0.12\pm 0.11$ &$-0.31\pm 0.21 $ & $-0.28\pm 0.07$ & 1.2 & 2.0  \\ 
717     & 20 & 5042  & 2.50 & 2.50 & $-0.08\pm 0.13$ & $+0.09\pm 0.11$ &$-0.25\pm 0.21 $ & $-0.18\pm 0.07$ & 1.3 & 1.3  \\ 
894     & 34 & 4876. & 2.50 & 2.50 & $-0.27\pm 0.20$ & $-0.04\pm 0.14$ &$-0.10\pm 0.16 $ & $-0.25\pm 0.11$ & 1.4 &1.4   \\ 
432     & 28 & 4818. & 2.50 & 2.30 & $-0.94\pm 0.13$ & $-0.83\pm 0.12$ &$-0.17\pm 0.23 $ & $-0.13\pm 0.07$ & 1.3 & 1.3   \\
635     & 19 & 4843. & 2.50 & 2.50 & $-0.35\pm 0.13$ & $-0.33\pm 0.12$ &$-0.20\pm 0.17 $ & $-0.26\pm 0.09$ & 1.4 & 1.8   \\
867     & 20 & 4892. & 2.50 & 2.50 & $-0.49\pm 0.15$ & $-0.57\pm 0.19$ &$-0.03\pm 0.18 $ & $-0.09\pm 0.16$ & 1.3 & 2.0   \\
927     & 43 & 4880. & 2.50 & 2.75 & $-0.06\pm 0.17$ & $-0.03\pm 0.15$ &$-0.20\pm 0.16 $ & $-0.25\pm 0.12$ & 1.4 & 1.2   \\
709     & 41 & 4930. & 2.50 & 2.50 & $-0.09\pm 0.16$ & $-0.02\pm 0.20$ &$-0.18\pm 0.24 $ & $-0.24\pm 0.11$ & 1.5 & 1.5   \\
\\
\hline
\\
\multispan{10}{$^a$ Star numbers are from \citet{Marconi} field 1, available
through CDS\hfill}\\ 
\multispan{10}{\phantom{Star}at {\tt cdsarc.u-strasbg.fr/pub/cats/J/A+A/330/453/sagit1.dat}\hfill}\\
\multispan{10}{$^b$ this is star  [BHM2000] 139  of \citet{B00}\hfill}\\
\multispan{10}{$^c$ this is star  [BHM2000] 143  of \citet{B00}\hfill}\\
\end{tabular}
\end{center}
\end{table*}

When degraded at  a  resolution R=15000 
the mean difference in derived metallicity
(in the sense UVES -- [R=15000]) is +0.013 dex
with a standard deviation of 0.15 dex.
The mean difference in [$\alpha$/Fe]
is --0.07 dex with a standard deviation of
0.08 dex.
The mean difference in $\xi$ is --0.07 \kms
with a standard deviation of 0.29 \kms.
These comparisons are illustrated in Fig. \ref{br7_20}.
This shows that the  resolution
of 15000 is still sufficient to provide reliable
abundances (see Fig.\ref{conf}).
 Since the actual resolution of Giraffe is higher
it is even better suited to be used with our method.

\subsection{Other elements}
\begin{figure}
\centering
\resizebox{\hsize}{!}{\includegraphics[clip=true]{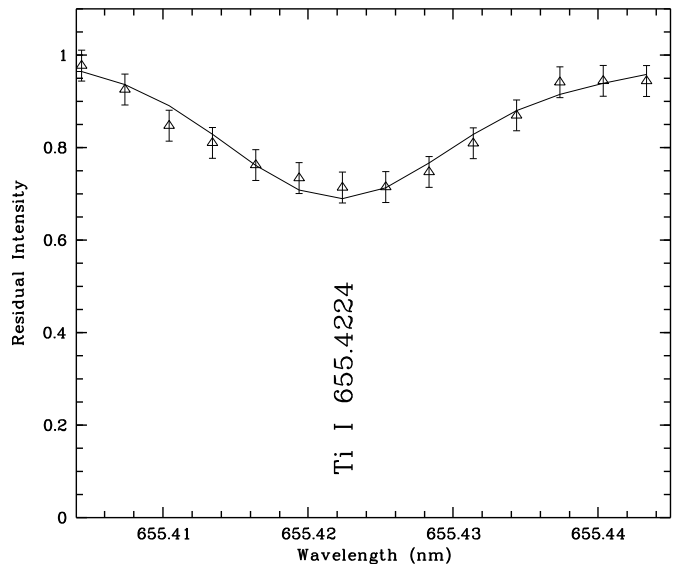}}
\caption{Plot of the fit of a Ti\,\textsc{i} feature in star \# 879.
Solid line is the fit done by the code; the triangles are the
experimental data pseudo-normalized; the bars are the errors
of the observed spectrum.}
\label{tiifit}
\end{figure}

 To check the behaviour of the code,
we analyzed a few other elements for the two
stars of \citet{B00}
for which a complete abundance analysis is available.
The results are reported in Table \ref{abboel}.
For the other stars of the sample,
the traditional 
abundance analysis is not yet complete.
As can be seen from the table
the results of  traditional and automatic analysis 
are compatible within $1\sigma $ (being $\sigma $
the dispersion of individual features when more than one
is available).
In Fig.\ref{tiifit} a fit for the Ti I 655.4224 nm line in star
\# 879 is plotted.
Implementation  of automatic analysis to obtain 
the abundances for several
elements is our goal in the next future, the present results
are encouraging.

\begin{table*} 
\caption{Results of automatic and traditional analysis for some elements}
\label{abboel}
\begin{center}
\begin{tabular}{lllllllll}
\hline
\\ 
Element &\hfill code\hfill  & $n$ &  traditional & $ n$ &\hfill code\hfill  & $n$   & traditional & $n$ \\
        & \multispan4{\hfill star \#879\hfill} & \multispan4{\hfill star \#772\hfill} \\
\\
\hline
\\
Ni\,\textsc{i} & $-0.64\pm 0.11$ & 6 & $-0.56\pm 0.16$ & 6 & $-0.44\pm 0.21$ & 6 & $-0.63\pm 0.10$ & 6\\
Si\,\textsc{i} & $-0.26\pm 0.12$ & 4 & $-0.35\pm 0.26$ & 5 & $-0.28\pm 0.09$ & 5 & $+0.06\pm 0.21$ & 4\\
Ti\,\textsc{i} & $-0.21\pm 0.05$ & 6 & $-0.19\pm 0.13$ & 7 & $-0.19\pm 0.13$ & 7 & $-0.21\pm 0.05$ & 6\\
Ti\,\textsc{ii}& $-0.47\pm 0.13$ & 2 & $-0.31\pm 0.07$ & 2 & $-0.46\pm 0.06$ & 2 & $-0.43\pm 0.18$ & 3\\
V\,\textsc{i}  & $-0.15        $ & 1 & $-0.21\pm 0.16$ & 2 & $-0.49        $ & 1 & $-0.44\pm 0.05$ & 2\\
\\
\hline
\end{tabular}
\end{center}
\end{table*}

\section { Conclusions }

We have built an automatic procedure to derive chemical
abundances from high resolution spectra.
The procedure is specifically targeted to the
analysis of giants of the Sgr dSph, thus the
grid of synthetic spectra and spectral features
employed in the analysis are optimized for stars
of spectral type and metallicity range appropriate
to this case.
Making use of Monte Carlo simulations we have shown that
the procedure is quite stable and provides highly
reproducible results.
Comparison with the results obtained on 12 UVES spectra
of Sgr dSph giants which have been independently analyzed
in the traditional way
by \citet{B00} and \citet{B02} allows us to conclude
that the automatic procedure provides results
which are consistent at $1 \sigma$ with the traditional
analysis.

We plan to use the procedure described in this paper
to obtain abundances for statistically significant
numbers of Sgr giants as soon as data from
the FLAMES facility are available. Our tests with
UVES spectra degraded at the resolution of 15000
suggest that both spectrographs fed by OzPoz shall
provide abundances of comparable accuracy, the lower
resolution of Giraffe being to some extent
compensated by the higher S/N achieved.

The reliability of our results is limited by
the grid of synthetic spectra used to interpret
the data. For example any error in the
atomic or molecular data used in the computation
will result in an error in the derived abundances.
However whenever new and better grids of synthetic
spectra are available it shall be straightforward
to implement them in the procedure.

One weak point of our procedure is to assume that
only the $\alpha$ elements have a direct
effect on the temperature structure of
the model.
Another is to assume that all the 
 $\alpha$ elements vary in lockstep and that
their ratio to iron is fixed by
the ratios of Mg and Ca.
The validity of this latter hypothesis may be checked directly by 
looking at the abundances of Si and Ti. 
If non solar ratios  of these elements to Mg and Ca are found the
results of our procedure may be called into question.

The concept of our procedure
may be generalized to the analysis of stars
of different spectral type and/or luminosity
class. In this case however it may not be enough
to compute a new set of synthetic spectra,  but also
the spectral features and spectral ranges
employed should be reconsidered based on the study
of real and synthetic spectra. 
  
\begin{acknowledgements}

We are grateful
to R. Cayrel for illuminating discussions
and suggestions on an early version of the
paper, as well as for his hospitality
at the Observatoire de Paris 
where much of this work was carried out.
We wish to thank also
B. Barbuy,  D. Katz and S. Zaggia
for their useful comments.
We thank the referee, P. North, for 
pointing out the weaknesses of the method 
and helping us in making
the paper clearer. 
This research was done with support from the
Italian MIUR COFIN2002 grant
``Stellar populations in the Local Group
as a tool to understand galaxy formation and evolution'' (P.I. M. Tosi).

\end{acknowledgements}

\bibliographystyle{aa}

\end{document}